\documentclass{cjour}
\usepackage{graphicx}

\def \xv {{\vec{x}}}
\def \kv {{\vec{k}}}
\def \Ih {{\hat{I}}}
\def \Kh {{\hat{K}}}
\def \Fh {{\hat{F}}}
\def \Comp {{\rm Comp}}
\def \Comm {{\rm Comm}}

\def \lim {{\rm lim}}

\def \BDM {\begin{displaymath}}
\def \EDM {\end{displaymath}}
\def \BEQ {\begin{equation}}
\def \EEQ {\end{equation}}
\def \BEQA {\begin{eqnarray}}
\def \EEQA {\end{eqnarray}}

\def \BL {\begin{list}}
\def \EL {\end{list}}
\def \BENUM {\begin{enumerate}}
\def \EENUM {\end{enumerate}}
\def \BITEM {\begin{itemize}}
\def \EITEM {\end{itemize}}
\def \BARR {\begin{array}}
\def \EARR {\end{array}}

\begin{document}

%%%%% To be entered at Academic Press: =====>>

% \journame{}
% \articlenumber{}
% \yearofpublication{}
% \volume{}
% \cccline{}
% \received{}
% \revised{}
% \accepted{}

% \authorrunninghead{}
% \titlerunninghead{}

% communication line, use: \commline{Communicated by...}
% \commline{ }

%\setcounter{page}{261} %% This command is optional. 

%% <<== End of commands to be entered at Academic Press

%%  Authors, start here ==>>

%\draft % Optional, will cause a line at the bottom of each page
%% with the words `Draft' and the time and date that the article
%% was LaTeXed. Will also double space text.

\title{A Multi-Threaded Fast Convolver for
       Dynamically Parallel Image Filtering
\thanks{
  This work is sponsored by DARPA, under Air Force Contract
F19628-00-C-0002.  Opinions, interpretations, conclusions and
recommendations are those of the author and are not necessarily
endorsed by the Department of Defense. 
}}

%\subtitle{}

\author{Jeremy Kepner}
\affil{MIT Lincoln Laboratory, 244 Wood St., Lexington, MA  02420}

%%%%%%%%%%%%
%% More than one author with separate affiliations, either:
%
%\author{First Author Name$^\dagger$ and Second Author Name$^\ddagger$}
%\affil{$^{\dagger}$First Author Affiliation, $^{\ddagger}Second Author
%       affiliation}

% or

% \author{Author name}
% \affil{Affiliation}
% \and
% \author{Author name}
% \affil{Affiliation}
%%%%%%%%%%%%

%% \thanks command:
%% Can use \thanks{} in title to have footnote number appear and
%% footnote at the bottom of the page. i.e.,
%% \title{This is the title\thanks{Supported by grant no....}}

%% In \authors or \affil, can use \thanks{} to have asterisk, 
%%   dagger or double dagger appear
%%   and text appear at the bottom of the title page. i.e.,

%\authors{D. Adalsteinsson and J. A. Sethian\thanks{Supported in part by the
%Applied Mathematics Subprogram of the...}}

%%%%%%%%%%%%

\email{kepner@ll.mit.edu}

%optional
%\dedication{Dedicated to...}

\abstract{
  2D convolution is a staple of digital image processing. The advent of
large format imagers makes it possible to literally ``pave'' with
silicon the focal plane of an optical sensor, which results in very
large images that can require a significant amount computation to
process. Filtering of large images via 2D convolutions is often
complicated by a variety of effects (e.g., non-uniformities found in
wide field of view instruments). This paper describes a fast (FFT based)
method for convolving images, which is also well suited to very large
images.  A parallel version of the method is implemented using a
multi-threaded approach, which allows more efficient load balancing and
a simpler software architecture.  The method has been implemented within
in a high level interpreted language (IDL), while also exploiting open
standards vector libraries (VSIPL) and open standards parallel
directives  (OpenMP).  The parallel approach and software architecture
are generally applicable to a variety of algorithms and has the
advantage of enabling users to obtain the convenience of an easy
operating environment while also delivering high performance using a
fully portable code.
}

% text should be lower case, unless caps are necessary for meaning

\begin{article}

% \contents is optional, will make a list of all section heads 
% that appear in the article
%\contents

%optional
% for those that like to start with section zero:
%\zerosection{Introduction}

% Here is the body of the article.

%%%%%%%%%%%%%%%%%%%%%%%%%%%%%%%%%%%%%%%%%%%%%%%%%%%%%%%%%%%%
\section{Introduction}
%%%%%%%%%%%%%%%%%%%%%%%%%%%%%%%%%%%%%%%%%%%%%%%%%%%%%%%%%%%%

  The ability to process ever larger images at ever increasing rates is
a key enabling technology for a wide variety of medical, scientific,
industrial and government applications (e.g., next generation 
MRI/x-ray, environmental monitoring, real-time digital video,
surveillance and tracking). Image filtering via 2D convolutions is often
the dominant image processing operation in terms of computation cost. 
Found in almost image processing pipelines 2D convolution is essential
for background compensation,  image enhancement, smoothing, detection
and estimation. In real-time imaging sensors rapid frame rates
require very low latency processing, which requires high performance 2D
convolutions.  In archival image database queries,  the size of the
image can be enormous and high performance convolutions are necessary in
order to complete the query in a reasonable amount of time.

  Traditionally, custom chips or coprocessors have been used to
alleviate image processing performance bottlenecks.  However, as image
processing applications become more complex and more software focused
(i.e. workstation based) these solutions become less feasible.  There
are essentially three ways of improving performance in a software
centered image processing environment: better algorithms, better
optimization and parallel processing.  This work applies all three of
these approaches specifically to improve the 2D convolution operation. 
Although the primary aim of this paper is to demonstrate a faster 2D
convolution algorithm, a secondary goal is to consider the system issues
necessary to effectively integrate this type of technology into readily
available image processing environments with minimal software
maintenance overhead.

  Section two of this paper presents a general FFT based algorithm for
implementing 2D convolutions, which is also well suited for wide field
of view images.  Section three describes a parallel scheme for
the convolution algorithm and provides general analyses the
computation, communication, and load balancing costs.  Section four,
discusses a general purpose software architecture for
implementing the algorithm and transparently integrating it within high
level image processing environments.  Section five presents the
parallel performance results.  Section six gives the
conclusions.

%%%%%%%%%%%%%%%%%%%%%%%%%%%%%%%%%%%%%%%%%%%%%%%%%%%%%%%%%%%%
\section{2D Filtering Algorithm}
%%%%%%%%%%%%%%%%%%%%%%%%%%%%%%%%%%%%%%%%%%%%%%%%%%%%%%%%%%%%

\begin{figure}
  \centerline{\includegraphics[width=6.5in]{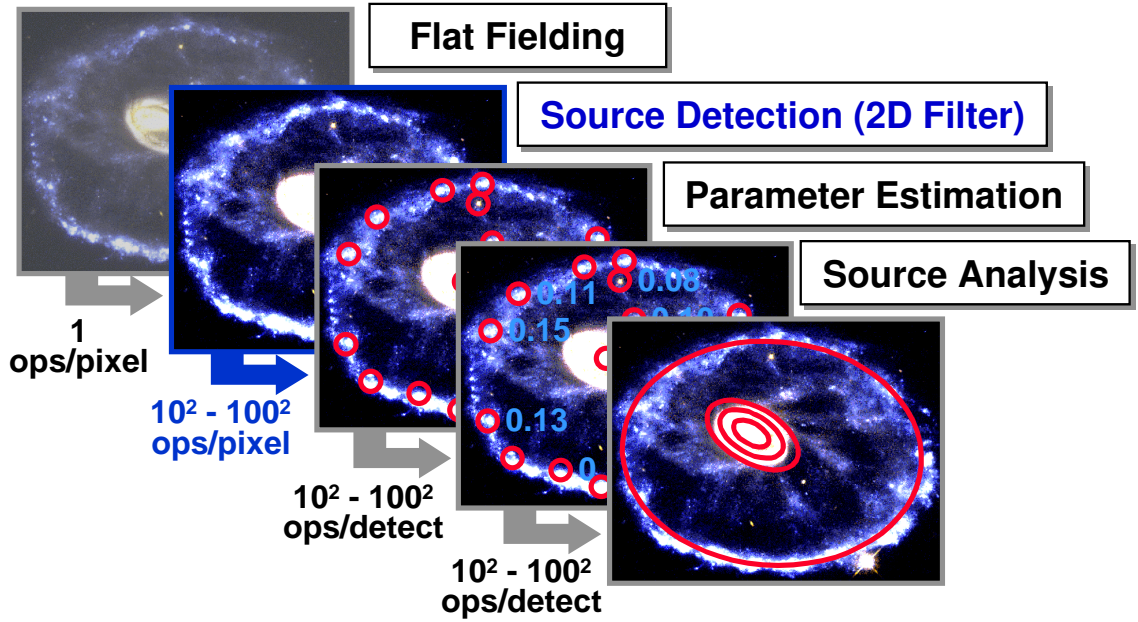}}
  \caption{{\bf Generic Image Processing Pipeline.}  Most pipelines
  consist of steps similar to the above.  In this process 2D filtering
  is the most compute intensive step.  Later steps are more complex
  (e.g. source analysis) but operate on a reduced amount of data.
  [Cartwheel Galaxy image courtesy of Kirk Borne (ST ScI), and NASA.]
  }
\label{fig:image_pipeline}
\end{figure}

  Wide area 2D convolution is a staple of digital image processing.
Figure~\ref{fig:image_pipeline} shows one example of how convolution
fits into an image processing pipeline.  Typically, an image $I(x,y)$
is acquired by a sensor or extracted from an archive.  We wish to
convolve or filter this image using a kernel $K(x,y)$ to produce a
filtered image $F(x,y)$.  Mathematically this is equivalent to
  \begin{equation}
     F(x,y) = \int \int K(x',y') I(x - x',y - y') dx' dy'
  \end{equation}
For convenience it is reasonable to assume the image and the kernel are
$M$ x $M$ and $N$ x $N$ squares, respectively (this condition can be
relaxed later without loss of generality).  In the discrete case, the
above convolution can be computed by the double sum
  \begin{equation}
     F(i,j) = \sum_{i'=0}^{N-1} \sum_{j'=0}^{N-1} K(i',j') I(i - i',j - j') ~ ,
  \end{equation}
where $i$ and $j$ are pixel indices. The above direct summation
involves $O(N^2)$ operations per pixel, which can become prohibitive
for large kernels.

\subsection{Basic FFT based algorithm}

\begin{figure}
  \centerline{\includegraphics[width=6.5in]{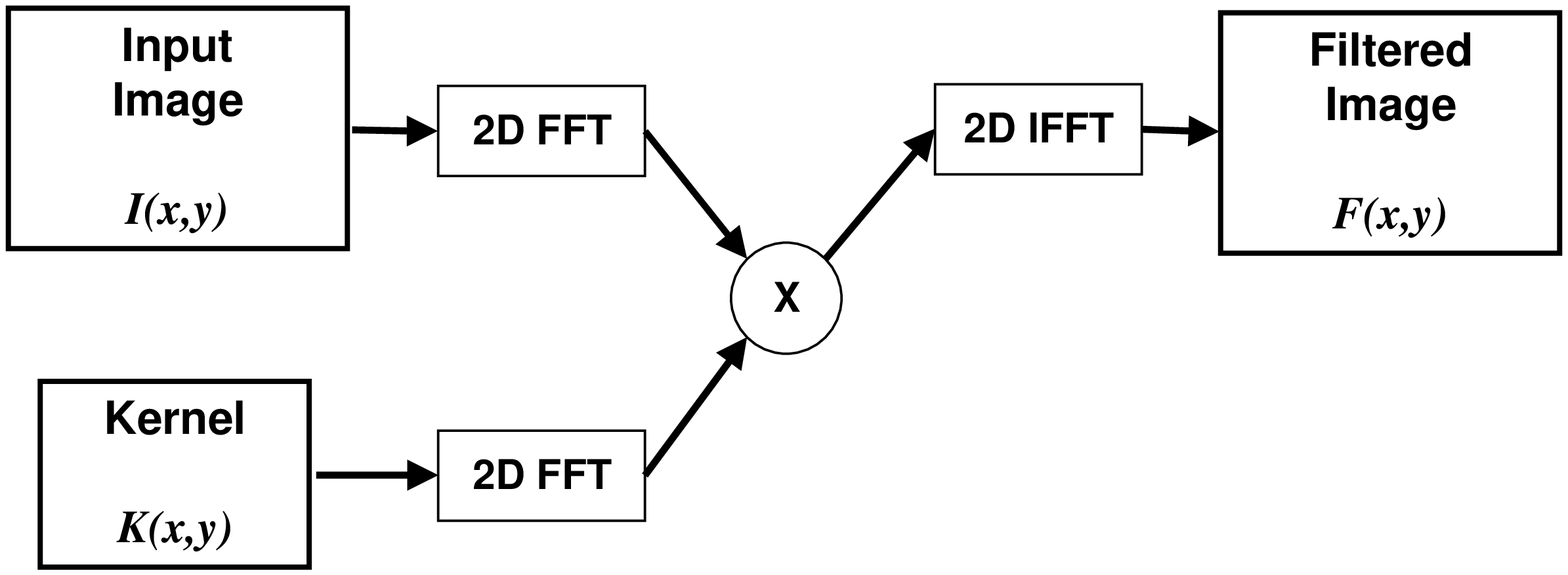}}
  \caption{{\bf Basic 2D Filtering.}  FFT implementation of
    2D filtering which performs the mathematical operation:
    $F(x,y) = \int \int K(x',y') I(x - x',y - y') dx' dy'$
  }
\label{fig:basic_2d_filtering}
\end{figure}

  A more efficient implementation of 2D filtering is obtained using
the Fast Fourier Transform.  This classic result exploits the fact that
convolution is equivalent to multiplication in the Fourier domain
  \begin{equation}
     \Fh(\kv) = \Kh(\kv) \cdot \Ih(\kv)
  \end{equation}
where $\Ih$, $\Kh$ and $\Fh$ are the Fourier Transforms
  \begin{eqnarray}
     \Ih(\kv) &=& \int I(\xv) e^{i \kv \cdot \xv} d\xv ~, \nonumber \\
     \Kh(\kv) &=& \int K(\xv) e^{i \kv \cdot \xv} d\xv ~, \nonumber \\
     \Fh(\kv) &=& \int F(\xv) e^{i \kv \cdot \xv} d\xv ~ ,
  \end{eqnarray}
where $\xv = (x,y)$ and $\kv = (k_x, k_y)$. Thus, in the discrete case
we can compute the convolution using the standard discrete 2D $FFT$ and
2D $FFT^{-1}$ functions (see Figure~\ref{fig:basic_2d_filtering})
  \begin{equation}
     F = FFT^{-1}(FFT(K)  FFT(I))
  \end{equation}
which has a computation cost of $O(\log_2(M+N)^2$) operations per
pixel, which, as will be shown later, can be reduced to
$O(\log_2 N^2)$.
[Traditionally, it has been necessary to pad the
image out to the nearest power of 2 in order to exploit optimized
implementations of the FFT.  This limitation can easily result in
``excessive'' padding in many image processing applications.  For
example, a 560 x 300 image may need to be padded out to 1024 x 512,
which is an increase of over 300\%.  However, more recently,
self-optimizating implementations of the FFT (e.g. FFTW
\cite{Frigo1998,Moler2001} allow high performance for non-powers of two
to be obtained without additional software tuning by programmers (see Appendix
A).]

\subsection{Convolution with Variable Kernels}
  As mentioned earlier, one of the primary drivers towards a more software
centered image processing pipelines is increased complexity. One example
of this type of complexity occurs in wide field of view imaging where
filtering of large images via 2D convolutions is often complicated by a
non-uniform point response function or kernel. For example, suppose the
kernel function is itself sampled over an $N_K$ x $N_K$ grid on the
image, so that $K_{ij}(x,y)$ represents the kernel centered on the $i,j$
point in the grid (Note: these indices are not the same as the pixel
indices used in the previous section).  The algorithm for dealing with
this case is well known and the particular approach used here is based
on the ``overlap-and-add'' method \cite{Stockham1966} for FFT based
filtering.  For discussion purposes the algorithm is first
presented for ``1D'' images and then the 2D algorithm is describe (the
method can be generalized to arbitrary dimensions, but that is not done
here).

\subsubsection{1D Case}
  Consider a 1D image to be filtered by a kernel which is sampled at
two points $K_1(x)$ and $K_2(x)$.  The true kernel at any point can be
approximated by the weighted average of the two
  \begin{equation}
    K(x') = W_1(x) K_1(x') + W_2(x) K_2(x')
  \end{equation}
where $W_1(x) = (x - x_1)/(x_2 - x_1)$ and $W_2(x) = 1 - W_1(x)$.  Thus,
filtering with this kernel becomes
  \begin{eqnarray}
    F(x) &=& \int K(x') I(x - x') dx' ~ , \nonumber \\
         &=& \int [W_1(x) K_1(x') + W_2(x) K_2(x')] I(x - x') dx' ~ , \nonumber \\
         &=& W_1(x) \int K_1(x') dx' ~~ + ~~ W_2(x) \int K_2(x') I(x - x') dx' ~ .
  \end{eqnarray}
In other words, convolving with a variable kernel is equivalent to
convolving with each kernel individually and combining the results with
appropriate weightings.

\subsubsection{2D Case}
\begin{figure}
  \centerline{\includegraphics[width=6.5in]{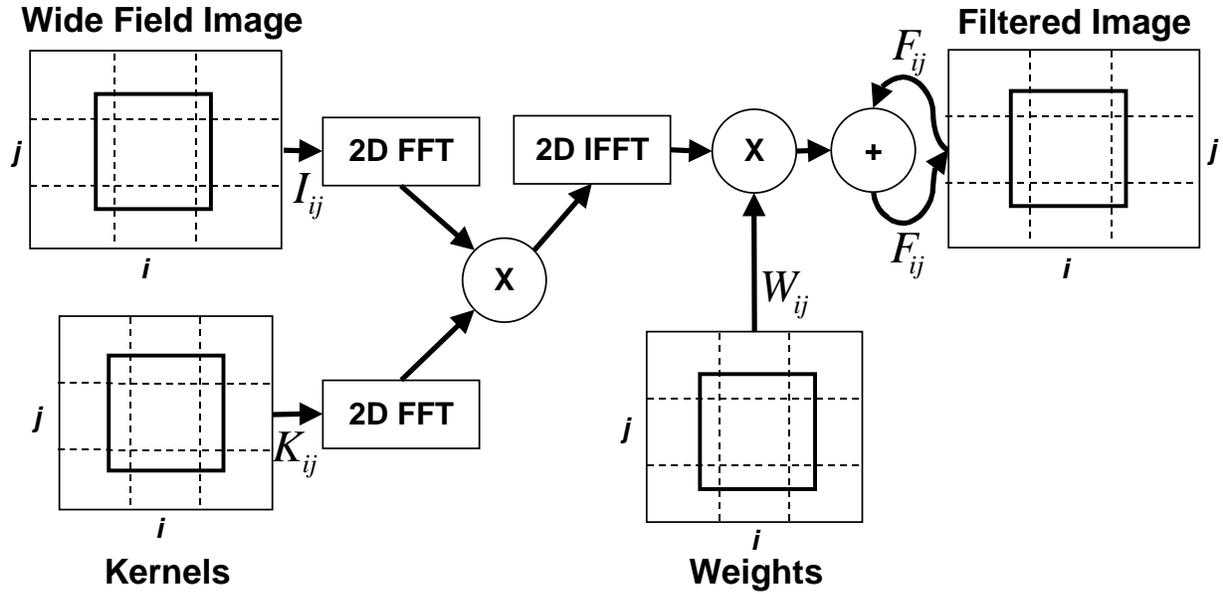}}
  \caption{{\bf Wide Field Filtering.}  FFT implementation of
    2D filtering for wide field imaging with multiple point response
    functions.  Each portion of image is filtered separately and
    then recombined using the appropriate weights.  The equivalent
    mathematical operation is:
    $F_{ij}(x,y) = W_{ij}(x,y) \int \int K_{ij}(x',y') I_{ij}(x - x',y - y') dx' dy'$
  }
\label{fig:wide_field_filtering}
\end{figure}

  2D images are a straightforward extension of the above case
  \begin{equation}
     F(x,y) = \sum_{i=0}^{N_K-1} \sum_{j=0}^{N_K-1} W_{ij}(x,y)
              \int \int K_{ij}(x,y) I(x - x',y - y') dx' dy' ~ .
  \end{equation}
Or, more explicitly, if $I_{ij}$ and $F_{ij}$ are the regions of $I$ and
$F$ that are affected by $K_{ij}$, then we can compute each $F_{ij}$
separately using FFT-based methods and then add it to the overall image
using the appropriate weightings (see
Figure~\ref{fig:wide_field_filtering}).

The above derivation is intended to be a rough outline of the classic
FFT based approach to 2D convolutions.  Many details have been left out
(e.g., real vs. complex FFTs, zero padding, treatment of edge effects,
etc ...).  For a more precise description of the details necessary to
implement the above algorithm see Appendix B.

%%%%%%%%%%%%%%%%%%%%%%%%%%%%%%%%%%%%%%%%%%%%%%%%%%%%%%%%%%%%
\section{Parallel Scheme}
%%%%%%%%%%%%%%%%%%%%%%%%%%%%%%%%%%%%%%%%%%%%%%%%%%%%%%%%%%%%

  The previous section presented a general algorithm for efficiently
convolving 2D images.  Up to this point, there has been no mention
as to how to implement this algorithm on a parallel computer.  There
are many potential ``degrees of parallelism'' available at the
coarse, medium and fine grain level.  These levels of parallelism
can be described as follows

\begin{description}
\item[{\bf Image}] The highest level of parallelism potentially exists
  at the application level.  It may be the case that multiple independent
  images are to be filtered in a sequence.  Exploiting this task level
  parallelism (sometimes referred to as ``round robining'') is very
  efficient, but is application specific and leads to long turnaround
  times for each image (i.e. high latency).
\item[{\bf Sub-image}] Convolving with multiple kernels (or breaking up a single
  convolution into multiple smaller convolutions) leads to a natural
  breakup of the larger image into sub-images.  This type of decomposition
  is sufficiently commonplace
  that it is reasonable to exploit this level of parallelism.  This level
  of parallelism is relatively coarse grain because the $I_{ij}$, $K_{ij}$,
  $W_{ij}$, and $F_{ij}$ can all be computed independently.  In addition,
  this level of parallelism can be abstracted away from the user.
\item[{\bf Row/Column}]  Within each image, it is possible to perform the convolutions
  first by row and then by column (or vice versa).  This offers a
  very large number of degrees of parallelism.  However, this method is
  complicated by the need to transpose or ``corner turn'' the data between
  steps, which can introduce large communication costs.  Furthermore, this
  method requires working beneath the 2D FFT routine, thus
  significantly increasing the coding overhead.
\item[{\bf Instruction}]  The lowest level of parallelism that can be exploited
  is at the instruction level.  This normally requires hardware support and is
  beyond the scope of this work.
\end{description}

  The selection of which level(s) of parallelism to employ is based on a
detailed analysis of the computation, communication, load balancing and
software overheads incurred.  These overheads are summarized in
Table~\ref{tab:parallel_overheads}.  Based on this analysis, exploiting
the natural decomposition of the larger image into sub-images was
selected. In this scheme each sub-image $I_{ij}$ is sent to a different
processor (see Figure~\ref{fig:wide_field_filtering}.  The approach has
a variety of advantages, not the least of which is that it can be
implemented within the scope of a math library and does not impose upon
the application (see next section). The rest of this section will
present a more detailed analysis of the overheads of this approach.
  
\subsection{Computation Cost}

  The computational cost of implementing an FFT based convolution is
dominated by the two forward and one inverse 2D FFTs (i.e. three FFTs
total).  In addition, each sub-image of the filtered image will be a
weighted sum of the convolution of the four neighboring kernels. The
cost per sub-image is $4\cdot 3\cdot 5 N'^2 \log_2 N'^2$, where $N' = N
+ M/N_K$ is the size each padded sub-image (Note: this padding is due to
edge effects and is unavoidable).  This cost can usually be reduced by
performing the FFT of the kernel in advance, and by exploiting the fact
that the images are real valued and not complex.  Using these
techniques, the computational cost can be reduced to $4\cdot 5 N'^2
\log_2 N'^2$.  The cost for computing the entire image is $N_K^2$ times
this value or $4\cdot 5 N_K^2 N'^2 \log_2 N'^2$.

   One advantage of the sub-image parallel approach is that it is
independent of the underlying method of performing the convolutions. A
direct summation approach can be used and may be more efficient in the
case of very small kernels.  If a direct summation approach is used,
this would require $4\cdot 2 N_K^2 N'^2 N^2$ operations.  The
``turnover'' point between direct summation and FFT based methods occurs
when $2 N^2 = 5 \log_2 N'^2$.  In the case where $N_K$ is large and
$M/N_K \approx N$, and $N' \approx 2 N$, then the turnover point occurs
when
  \begin{equation}
      4\cdot 5 N_K^2 N'^2 \log_2 N'^2 = 4\cdot 2 N_K^2 N'^2 N^2
  \end{equation}
or
  \begin{equation}
     N^2 = 5 \log_2(4N).
  \end{equation}
In other words, if $N > 8$ then it is more efficient to
use an FFT based approach.  [This estimate is only
a guideline as many other system specific factors (cache architecture,
vector size, pipeline depth, etc ...) effect the precise performance
of these operations.]

\subsection{Communication Cost}

  The impact of communication depends upon the distributions of
the input and output images and the memory and networking architecture
of the parallel computer.

  In the expected case, communication costs of this approach are
dominated by the initial sending of data to each processor and the final
assembly of the filtered sub-images into one image.  Because of edge
effects, slightly more than the entire image needs to be sent to the
individual processors, $N'^2$ pixels per sub-image or $N_K^2 N'^2 = M^2
+ N_K^2 N^2$ pixels in total.  To assemble the final image, there are no
edge effects, but requires summing up to four separate sub-images or $4
M^2$ pixels total.

  In the best case, if each sub-image is initially distributed onto a
processor the initial sending of data reduces to sending the
overlapping edge data or $N_K^2 N^2$ pixels, and the final assembling
of data will be reduced to $3 M^2$ pixels (not a large change from the
expected case).

  In the worst case, all of the data reside on a single processor,
which doesn't change the amount of data transmitted (as compared to the
expected case), but can lead to a bottleneck due to the finite link
bandwidth into and out of a single processor.

  These communication costs, when combined with computation costs
leads to a computation to communication ratio of
\begin{equation}
  R = \frac{{\rm Computation}}{{\rm Communication}} =
  \frac{20 N_K^2 N'^2 \log_2 N'^2}{N_K^2 N'^2 + 4 M^2} .
  \label{eq:comp_comm}
\end{equation}
In the case where $N_K$ is large and $M/N_K \approx N$,
and $N' \approx 2 N$ then $M \approx \frac{1}{2} N_K N'$, and
\begin{equation}
  R \approx
  \frac{20 N_K^2 N'^2 \log_2 N'^2}{N_K^2 N'^2 + N_K^2 N'^2} =
  10 \log_2 N'^2 ~, 
\end{equation}
which for a typical value (e.g., $N' = 256$) leads to a computation
to communication ratio of 160, which indicates that reasonably good
parallel speedups should be possible with this parallel scheme.

  More specifically, the parallel speedup can be roughly modeled as
\begin{equation}
  {\rm Speedup} = \frac{T_\Comp}{T_\Comp/N_P + T_\Comm}
                = \frac{(T_\Comp/T_\Comm)}{(T_\Comp/T_\Comm)/N_P + 1}
                = \frac{\alpha^{-1} R}{\alpha^{-1} R/N_P + 1}
  \label{eq:speedup}
\end{equation}
where $N_P$ is the number processors, $T_\Comp/T_\Comm = \alpha^{-1} R$,
and $\alpha$ parameterizes the performance of the parallel computer
\begin{equation}
  \alpha = \frac{\rm Efffective~Processor~Speed}{\rm Effective~Link~Bandwidth}
         \approx \frac{0.15 ({\rm Peak~Speed})}{0.5 ({\rm Peak~Bandwidth})} ~.
\end{equation}
For a typical loosely connected parallel computer such as a cluster
$\alpha \approx 3$ flop/byte.  In such a system, the maximum speedup
obtainable ($N_P \rightarrow \infty$) is $\alpha^{-1} R \approx 50$. 
For a more tightly coupled system $\alpha \approx 1$ and the maximum
speedup is closer to 150.

\subsection{Load Balancing}

  Load balancing is critical to the effective use of a parallel system.
Finite images with edges introduce load imbalances.  The four corner
sub-images will have to do only one quarter of the processing and the $4
(N_K - 2)$ edge sub-images will have roughly one half the processing of the
$(N_K - 2)^2$ interior sub-images.  In addition, if the image is not
square additional imbalances will occur.  Depending on the
distribution of the data, the amount of time it will take to communicate
to different processors depends on their location within the computer,
which can lead to further imbalances.

\begin{figure}
  \centerline{\includegraphics[width=6.5in]{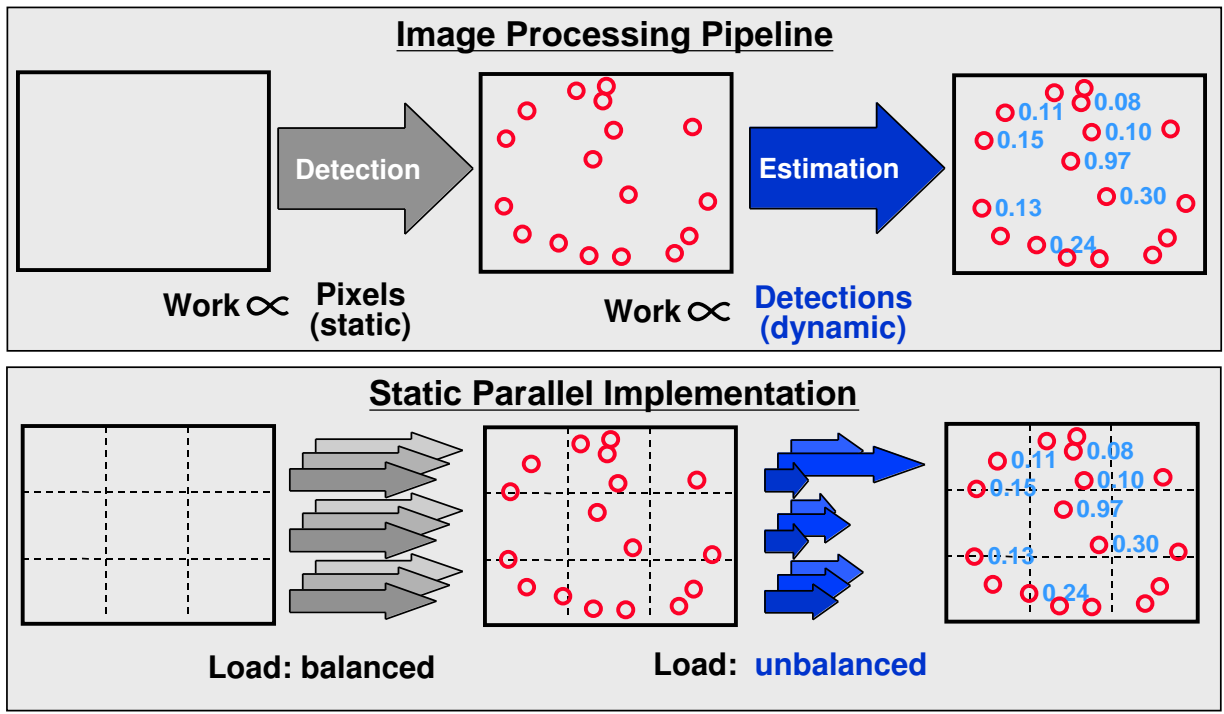}}
  \caption{{\bf Load Imbalance.}  Pre-detection the work in a
  pipeline is usually a simple function of the number of pixels.  Post-detection
  the work is usually a simple function of the number of detections.
  In a statically parallel system, statistical fluctuations in
  the number of detections will lead to a load imbalance in the
  post detection processing.
  }
\label{fig:load_imbalance}
\end{figure}

  In addressing these load balancing issues, it is worthwhile to
consider how filtering is used within the overall image processing
chain (see Figure~\ref{fig:load_imbalance}). It is often the case that filtering
immediately precedes the detection stage in an image processing
pipeline, which marks the boundary between the deterministic and
stochastic processing loads. Before detection, the amount of processing
is simply a function of the size of the input image and can be
determined in advance. After detection, the work load is proportional
to the number of detections which are randomly distributed.

  Randomly distributed loads are a challenging and well studied problem
\cite{Shirazi1995}. In general, dynamic load balancing is required in
order to effectively handle these types of problems (see Appendix C). 
Dynamical load balancing approaches often use a central authority to
assign work to processors as they become available. Thus, from a higher
level system perspective, given the need for dynamic load balancing
approaches in the system, it is worthwhile to exploit them to handle the
deterministic load imbalance introduced by edge effects.

  The parallel algorithm employed here lends itself to dynamic load
balancing for two reasons. First, the input data is never modified and
can be broadcast in any order.  Second, the final assembly
consists of weighted sums which can also be performed in any order. 
The ability perform operations out of order allows a great deal of
flexibility in pursuing dynamic load balancing techniques.  Of course,
the price for employing these mechanisms is the need for the underlying
hardware and software technologies that support them, such as shared
memory and multi-threading.

\subsection{Software Costs}

  As mentioned earlier, one of the primary benefits of parallelizing
over the sub-images is that it allows all the parallel complexity be
implemented in a way that is hidden from the user.  Thus this parallel
convolution function can be implemented with a very simple signature

\begin{verbatim}
       filtered_image = convolve(input_image,kernels,n_processors)
\end{verbatim}

Such a lightweight application level signature places all of the burden
of the implementation into the library, but this burden can be
significantly eased by employing existing open standards for doing high
performance mathematical operations and thread based parallelism (see
next section)

%%%%%%%%%%%%%%%%%%%%%%%%%%%%%%%%%%%%%%%%%%%%%%%%%%%%%%%%%%%%
\section{Software Architecture}
%%%%%%%%%%%%%%%%%%%%%%%%%%%%%%%%%%%%%%%%%%%%%%%%%%%%%%%%%%%%

  The previous sections presented and analyzed an algorithm and
parallel scheme for image filtering operations.  The analysis indicates
that significant speedups should be possible with this approach.
However, this potential will be of limited value unless it can
be incorporated into an effective software architecture, which addresses
the issues of portability, performance and productivity.  This section
describes how these issues are addressed for the 2D convolution
algorithm, using software techniques that are generally applicable
to a wide variety of algorithms.

\begin{figure}
  \centerline{\includegraphics[width=6.5in]{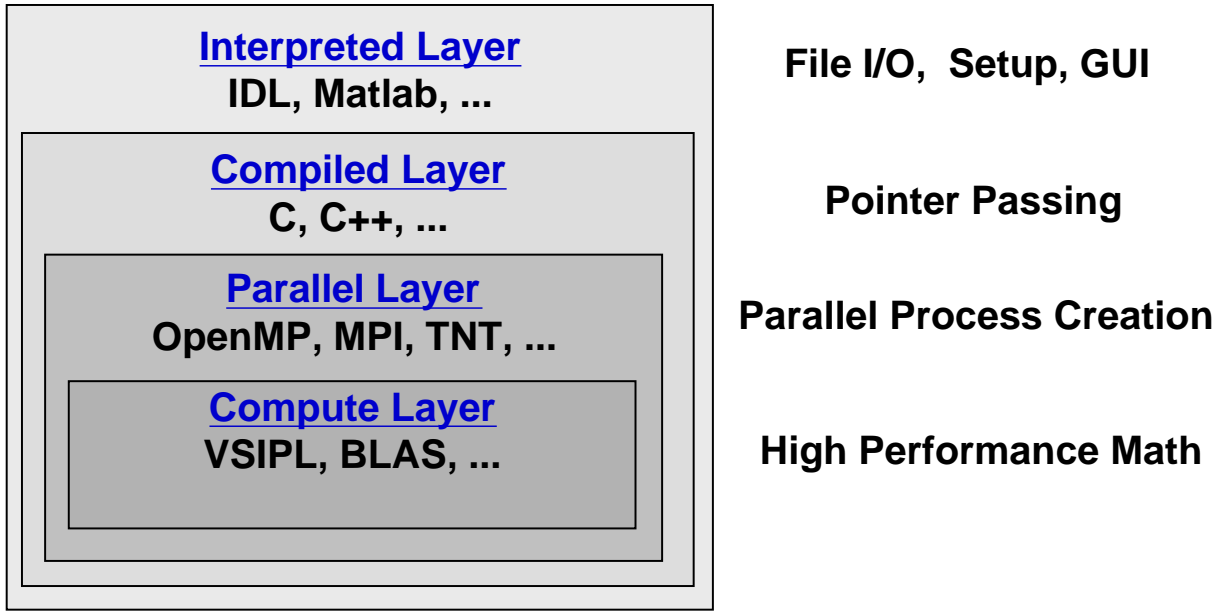}}
  \caption{{\bf Layered Software Architecture.}  The user interacts
    with the top layer which provides high level abstractions for
    high productivity.  Lower layers provide performance via parallel
    processing and high performance kernels.
  }
\label{fig:software_layers}
\end{figure}

  The key to supporting a parallel algorithm with effective software is
to use a layered approach with each layer addressing a different system
requirement (see Figure \ref{fig:software_layers}).  At the top level a
high level interpreted language (such as IDL or Matlab) is used.  This
allows for rapid integration of the parallel 2D convolver into a high
productivity environment by simply adding a function call to these
environments. The next layer is the parallel library layer (in this
case OpenMP) which provides access to thread-based parallelism using an
open portable standard.  The lowest level is the computation layer
where the actual mathematical operations are performed.  Here the 
Vector, Signal, and Image Processing Library (VSIPL) standard is used
to provide high performance computations using an open standard. Both
the OpenMP and VSIPL standards have enormous potential to allow users
to realize the goal of portable applications that are both parallel and
optimized.

  VSIPL is an open standard C language Application Programmer Interface
(API) that allows portable and optimized single processor programs. This
standard encompasses many core mathematical functions including FFT and
other signal processing operations which are essential for 2D filtering.
In addition, VSIPL provides strong support for early binding and in
place operations which allows the overhead of setup and memory
allocation to be dealt with outside of the time critical part of the
program.

\begin{figure}
  \centerline{\includegraphics[width=6.5in]{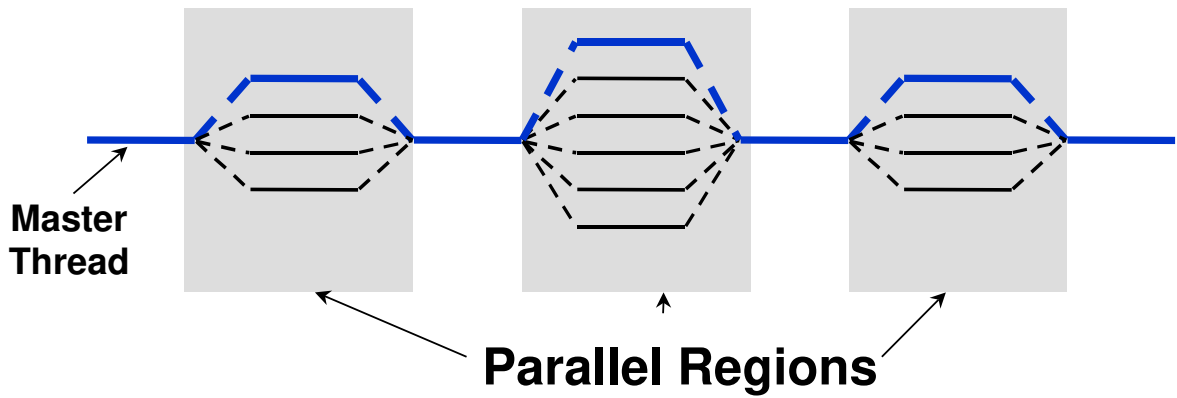}}
  \caption{{\bf Multi-threaded Program.}  OpenMP uses a fork/join
  model in which a master thread forks off multiple parallel threads,
  which rejoin to communicate or synchronize (figure adapted from \cite{Matson1999})
  }
\label{fig:program_control}
\end{figure}

   OpenMP \cite{OpenMP} is an open standard C/Fortran API that allows
portable thread based parallelism on shared memory computers.  OpenMP
uses a basic fork/join model (see Figure~\ref{fig:program_control})
wherein a master thread forks off threads which can be executed in
parallel and rejoins them when communication or synchronization is
required.  This very simple model allows for a large program to be
parallelize quickly with the insertion of a small number of compiler
pragmas.  As described earlier thread based parallelism is advantageous
for two reasons.  First, it is highly amenable to dynamic load balancing
schemes.  Second, it is easily implemented underneath higher level
environments because it based on dynamic process creation and does not
require a priori process creation or multiple invocations of the higher
level library.

  Exploiting these new open standards requires integrating them into
existing applications as well as using them in new efforts.  Image
processing is one of the key areas where VSIPL and OpenMP can make a
large impact.  Currently, a large fraction of image processing
applications are written in the Interpreted Data Language (IDL)
environment \cite{IDL}.  A goal of this work is to show that it is possible to
bring the performance benefits of these new standards to the image
processing community in a high level manner that is transparent to
the user.  IDL, like most interpreted languages, does not have parallel
constructs, but has a simple means for linking to externally built
library functions.  This mechanism allows the user to write
functions in Fortran, C or C++ to obtain better performance.  In addition,
it is possible to link in other high performance and parallel libraries
in a manner that is independent of the calling mechanism.

  There are many opportunities for parallelism in this algorithm. The
one chosen here is to convolve each kernel separately on a different
processor and then combine all the results on a single processor.  At
the top level a user passes the inputs into an IDL routine which passes
pointers to an external C function.  Within the C function OpenMP forks
off multiple threads.  Each thread executes its convolution using VSIPL
functions.  The OpenMP threads are then rejoined and the results are
added.  Finally a pointer to the output image is returned to the IDL
environment in the same manner as any other IDL routine.

%%%%%%%%%%%%%%%%%%%%%%%%%%%%%%%%%%%%%%%%%%%%%%%%%%%%%%%%%%%%
\section{Results}
%%%%%%%%%%%%%%%%%%%%%%%%%%%%%%%%%%%%%%%%%%%%%%%%%%%%%%%%%%%%

  The inputs of image convolution with variable kernels consists of a
source image, a set of kernel images, and a grid which locates the
center of each kernel on the source image.  The output image is the
convolution of the input image with each PRF linearly weighted by its
distance from its grid center.  Today, typical images sizes are in the
millions (2K x 2K) to billions (40K x 40K) of pixels.  A single kernel
is typically thousands of pixels (100 x 100) pixels, but can be as
small 10 x 10 or as large as the entire image.  Over a single image a
kernel will be sampled as few as once but as many as hundreds of times
depending on the optical system.

  The parallel convolution algorithm presented here was implemented on
an SGI Origin 2000 at Boston University.  This machine consists of 64
300 MHz MIPS 10000 processors with an aggregate memory of 16 GBytes. 
IDL version 5.3 from Research Systems, Inc. was used along with SGI's
native OpenMP compiler (version 7.3.1) and the TASP VSIPL
implementation. Implementing the components of the system was the same
as if each were done separately.  Integrating the pieces
(IDL/OpenMP/VSIPL) was done quickly, although care must be taken to use
the latest versions of the compilers and libraries.  Once implemented
the software can be quickly ported via Makefile modifications to any
system that has IDL, OpenMP, and VSIPL (currently these are SGI, HP,
Sun, IBM, and Red Hat Linux). 

  The performance of this implementation was tested by timing the
program over a variety of inputs and numbers of processors. The measured
speedups were computed by dividing the parallel times by the single
processor times (see Table~\ref{tab:parallel_speedups} and
Figure~\ref{fig:parallel_speedup}). In all cases, the kernel size was $N
= 100$ and the padded sub-image size was $N' = 512$.  The resulting
computation to communication ratio obtained from
Equation~\ref{eq:comp_comm} was $R = 288$.  Assuming $\alpha \approx
1.0$ and inserting this value of $R$ in Equation~\ref{eq:speedup} gives
predicted speedups of 2.0,  3.9, 7.8, 15.2, and 28.8 on 1, 2, 4, 8, 16
and 32 processors, respectively The measured and predicted parallel
efficiencies (speedup/number of processors) are shown in
Figure~\ref{fig:parallel_efficiency}, and are good agreement.

%%%%%%%%%%%%%%%%%%%%%%%%%%%%%%%%%%%%%%%%%%%%%%%%%%%%%%%%%%%%
\section{Conclusions}
%%%%%%%%%%%%%%%%%%%%%%%%%%%%%%%%%%%%%%%%%%%%%%%%%%%%%%%%%%%%

\begin{figure}
  \centerline{\includegraphics[width=6.5in]{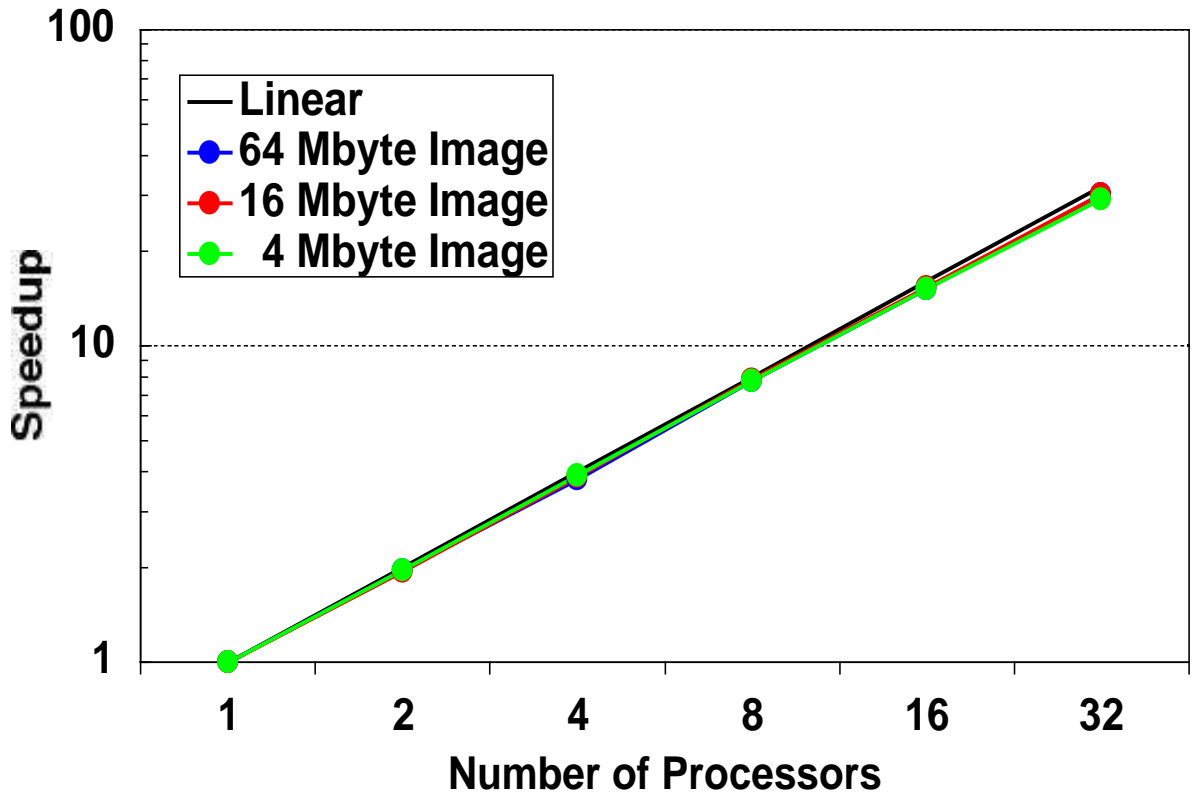}}
  \caption{{\bf Parallel Speedup.}  Measured speedups of
     wide field 2D filtering application on an shared memory
     parallel system (SGI Origin2000).
  }
\label{fig:parallel_speedup}
\end{figure}

\begin{figure}
  \centerline{\includegraphics[width=6.5in]{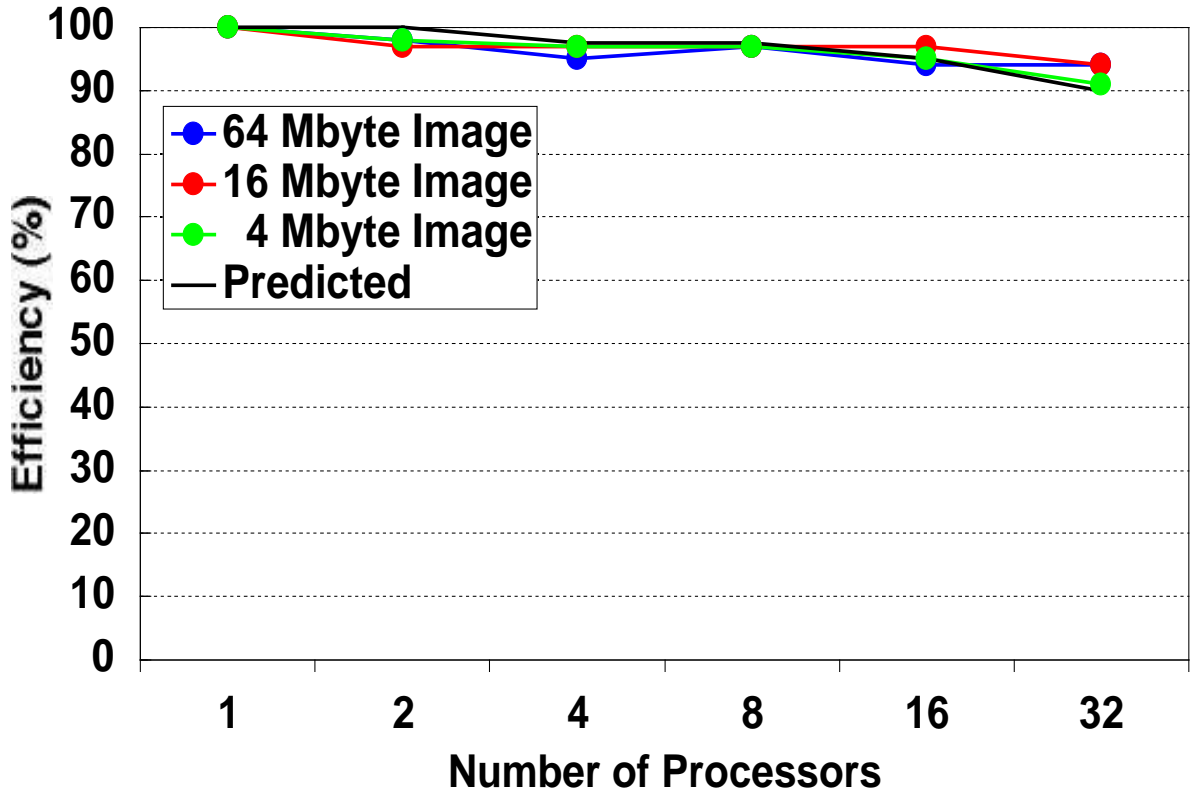}}
  \caption{{\bf Parallel Efficiency.}  Measured parallel efficiency of
     wide field 2D filtering application on an shared memory
     parallel system (SGI Origin2000).
  }
\label{fig:parallel_efficiency}
\end{figure}

  Image filtering via 2D convolutions is often the dominant image
processing operation in terms of computational cost.  This work has
looked at three ways of improving performance of 2D convolutions in a
software centered image processing environment: better algorithms,
better optimization and parallel processing. The result is a general FFT
based algorithm for implementing 2D convolutions, which is also well
suited for wide field of view images.  The algorithm uses a parallel
scheme which minimizes the communication overhead and allows for dynamic
load balancing. The algorithm has been implemented using a general
purpose  software architecture for transparently integrating it within
high level image processing environments.  This implementation exploits
the OpenMP and VSIPL standards. We have conducted a variety of
experiments which show linear speedups using different numbers of
processors and different image sizes (see
Figures~\ref{fig:parallel_speedup} and \ref{fig:parallel_efficiency}). 
Thus, it is reasonable to conclude that it is possible to achieve good
parallel performance using open standards underneath existing high level
languages.

  This general approach is not limited to IDL but can be extended
to most interpreted languages (e.g., Matlab, Mathematica, ...).
Extending this approach to other environments and implementing
a variety of important signal processing kernels (e.g. pulse compression,
adaptive beam-forming, ...) has enormous potential for enabling
easy to use high performance parallel computing.

%% End of article:

%% optional
\begin{acknowledgment}
The author is grateful to Glenn Breshnehan, Kadin Tsang and Mike Dugan
of the Boston University Supercomputer Center for their assistance in
implementing the software; to Randy Judd and James Lebak for their
assistance with the VSIP Library; to Charlie Rader for helpful comments
on 2D FFTs; to Matteo Frigo and Steve Johnson for their assistance with
FFTW; to Paul Monticciolo for his help with the text and to Bob Bond for
his overall guidance.
\end{acknowledgment}

% Appendix with letter and title:
\appendix{A}
\appendixtitle{Optimal padding of FFTs}

  Parallel implementations of any algorithm need reliable models for
the underlying computations.  The standard computational cost of a
complex FFT of length $N$ is $5 N \log_2(N)$.  The FFT can be
implemented with a variety radixes, but normally base 2 is the simplest
to optimize.  In a typical optimized FFT implementation powers of
two FFT sizes usually offer a significant performance boost over other
sizes.  Furthermore, because of the butterfly data access patterns of the
FFT, the performance penalty of using a non-optimized FFT can be quite
large (especially on a computer with a multi-level cache).  Given this
situation, it has been standard practice to ``pad'' the FFTs out to
the nearest power of two $N' = 2^{\lceil \log_2(N) \rceil}$.

\begin{figure}
  \centerline{\includegraphics[width=6.5in]{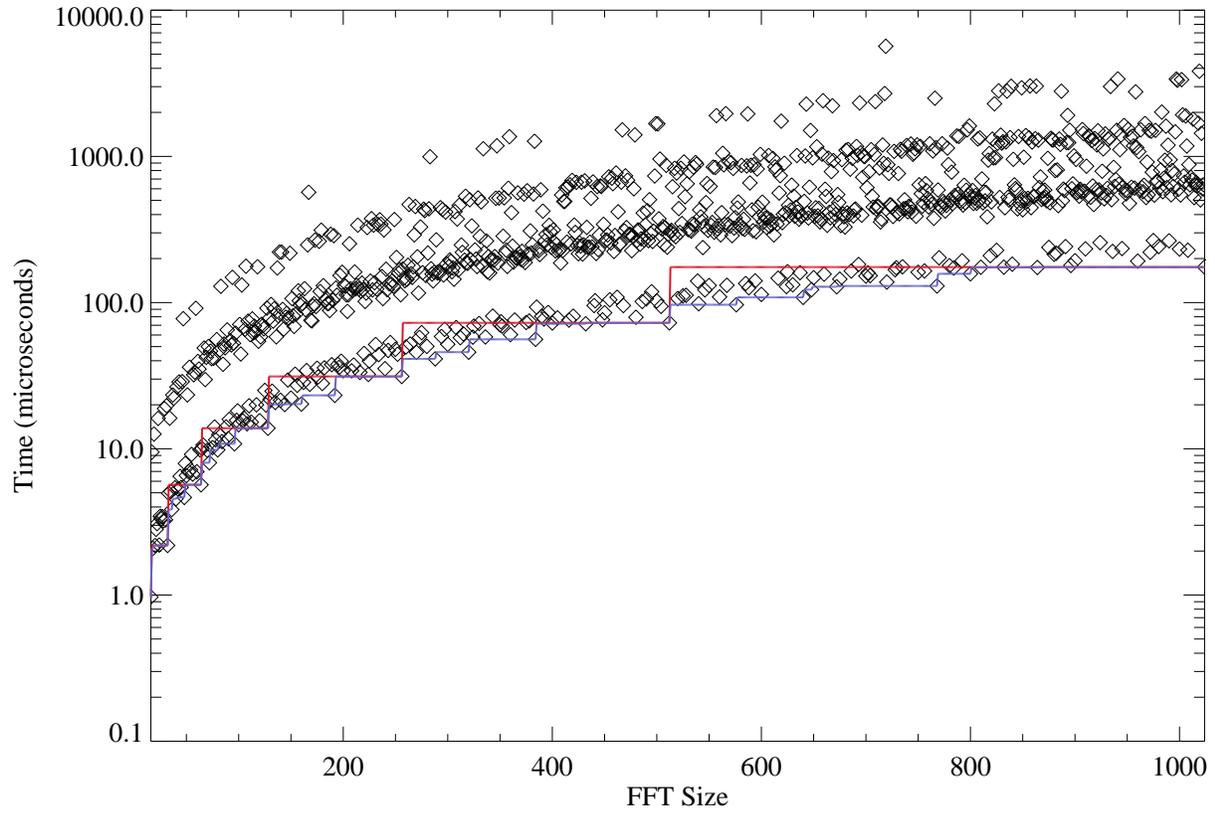}}
  \caption{{\bf FFTW performance.}  Performance of FFTW on
  FFTs ranging in size from 16 to 1024.  The blue line indicates
  the performance achieved using the optimally padded FFT.  The
  red line shows the performance obtained using powers of two.
  }
\label{fig:fft_timings}
\end{figure}

\begin{figure}
  \centerline{\includegraphics[width=6.5in]{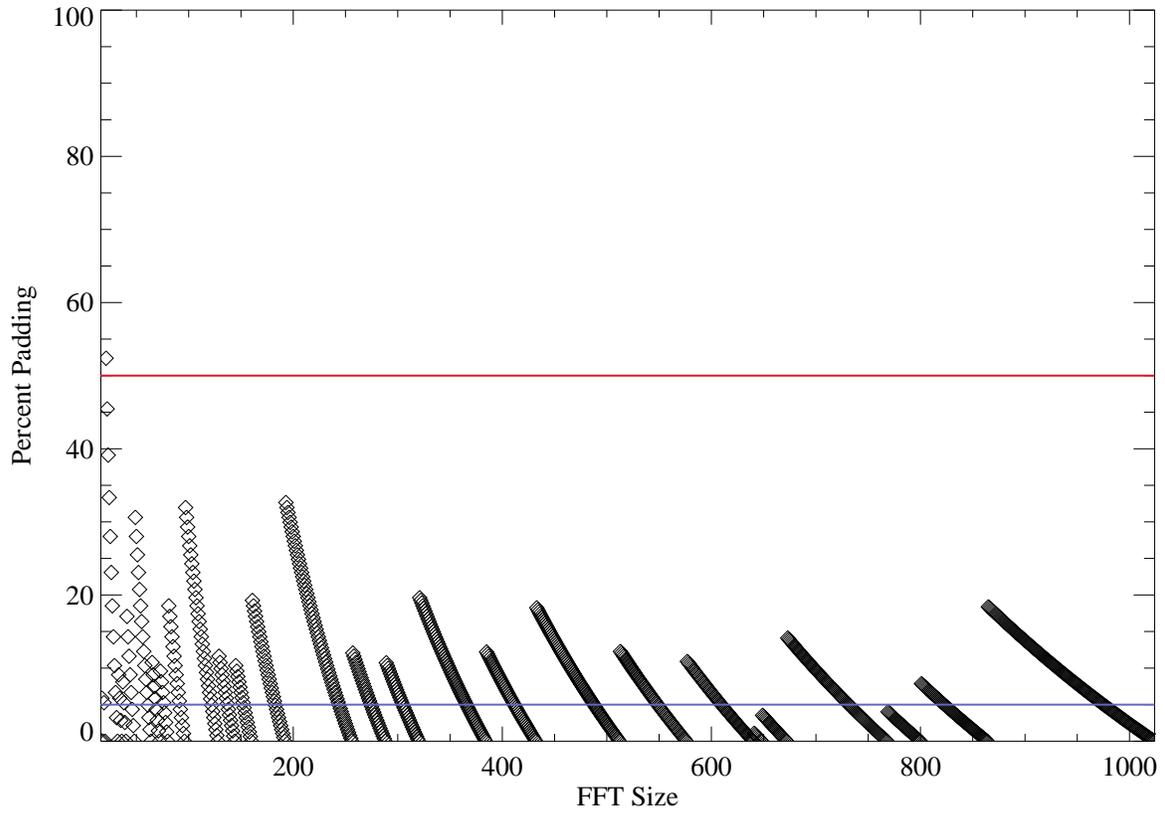}}
  \caption{{\bf FFTW Optimal Paddings.} Optimal padding (as a percentage of
  FFT size).  The ability to get good performance with non-powers of two
  significantly reduces the amount of padding required.
  The blue line indicates the average padding require across all the
  different cases.  The red line shows the average padding required
  if only powers of two are used.
  }
\label{fig:fft_paddings}
\end{figure}

\begin{figure}
  \centerline{\includegraphics[width=6.5in]{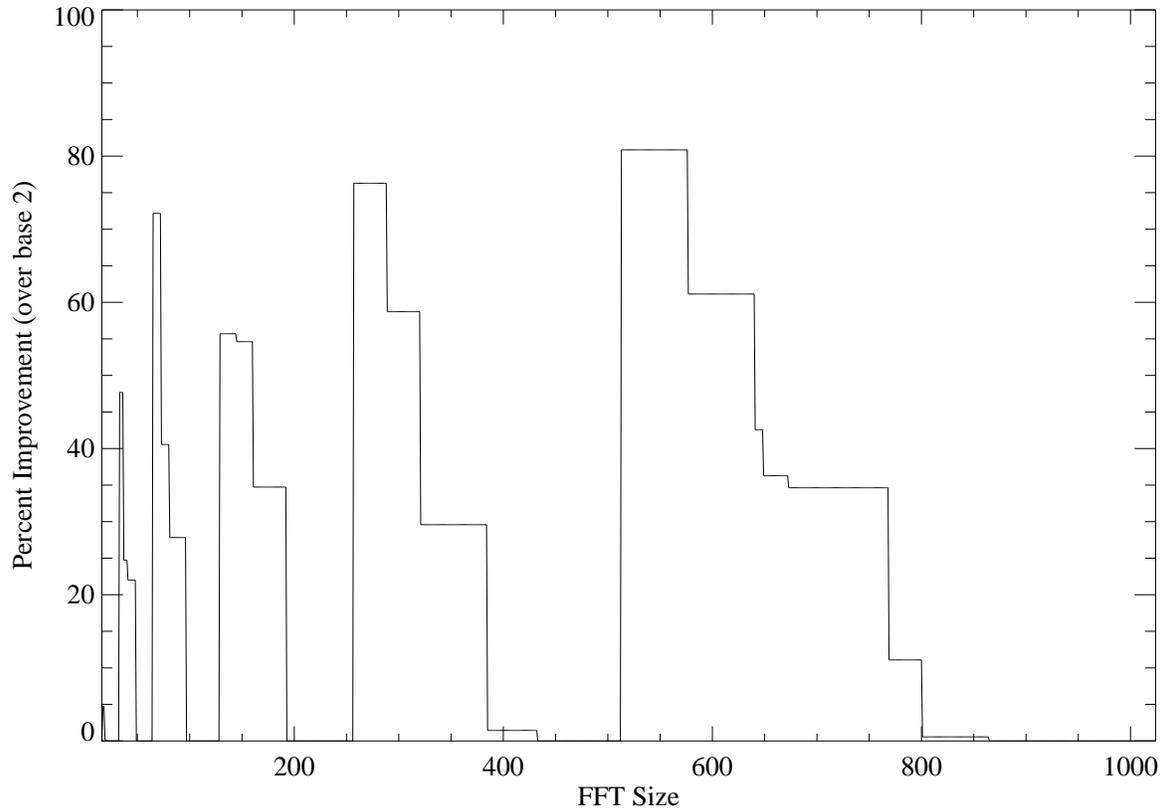}}
  \caption{{\bf FFTW Performance Benefit.}  The percentage performance
  improvement obtained using optimal padding compared to only powers of
  two.
  }
\label{fig:fft_base2}
\end{figure}

  Fortunately, the advent of self-optimizing software libraries such as
FFTW now significantly reduce the software engineering cost of
optimizing non-standard length FFTs.  Figure~\ref{fig:fft_timings} shows the
times for different length FFTs using FFTW, and the results of using the
optimal amount of padding.  The amount of padding and the performance
gains of using this padding are shown in Figure~\ref{fig:fft_paddings}.  
The performance gained using a non-powers of two FFT are shown in
Figure~\ref{fig:fft_base2}.  These results indicate that while some padding is
still a good idea, it is usually quite modest ($\sim$5\%).  Given this
situation it is now quite reasonable to estimate FFT performance using
unpadded values.

% Appendix with letter and title:
\appendix{B}
\appendixtitle{2D Convolution Algorithm Description}

\begin{figure}
  \centerline{\includegraphics[width=6.5in]{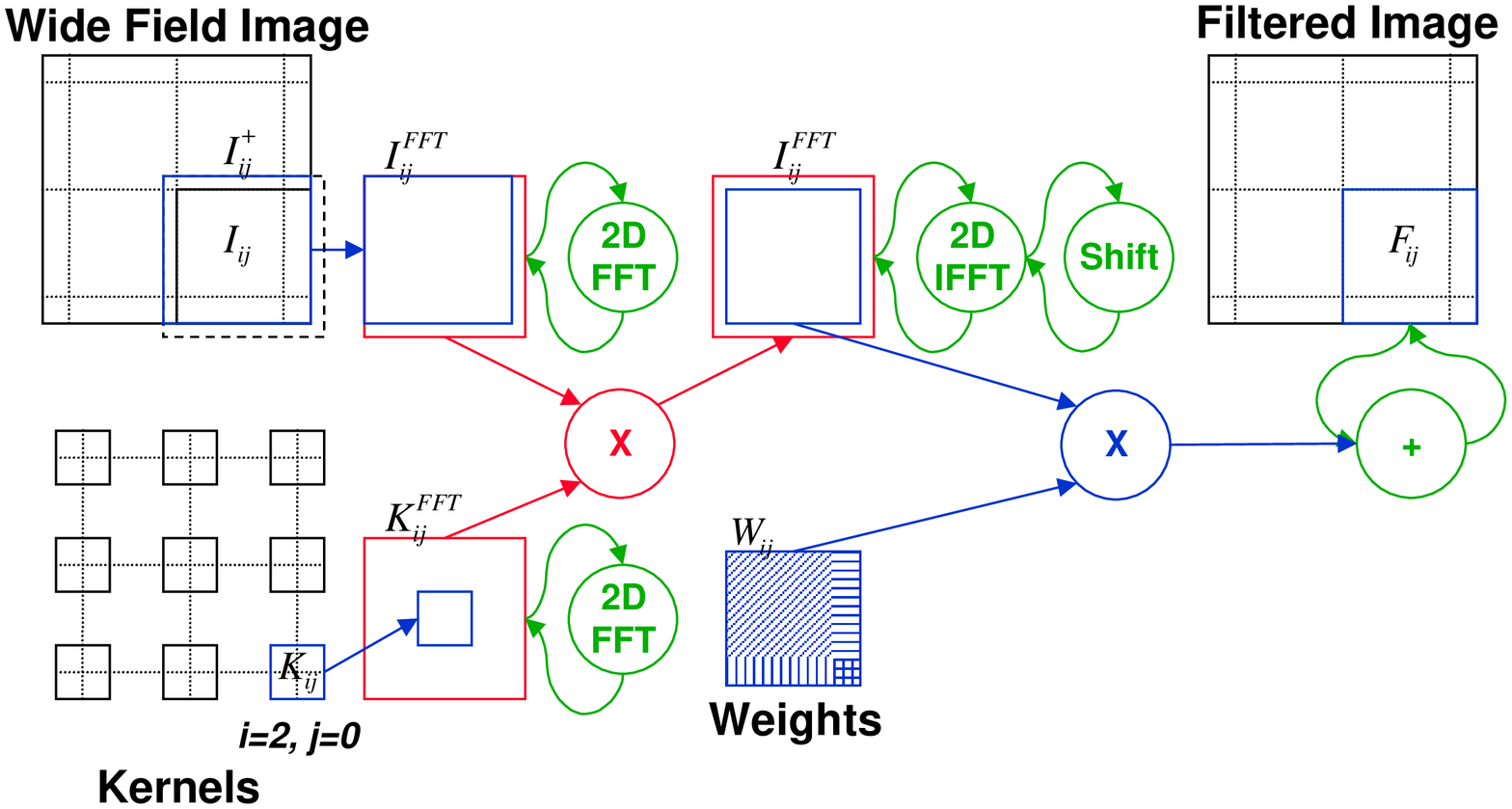}}
  \caption{{\bf Detailed Algorithm Flow.}  Precise computations,
    data structures and data movements necessary
    to execute 2D convolution algorithm with multiple kernels.
    The inputs consist of the image to be filtered and a grid with
    a kernel at each node.  The algorithm proceeds by looping 
    (in parallel) over each
    kernel in the grid $K_{ij}$ and executing the following steps: 
\begin{itemize}
  \item  Determine boundaries of input sub-image $I_{ij}$.
  \item  Compute corresponding weights $W_{ij}$.
  \item  Determine boundaries of filtered sub-image $F_{ij}$.
  \item  Determine boundaries of sub-image padded by kernel $I^{+}_{ij}$.
  \item  Determine size of and create padded sub-image $I^{FFT}_{ij}$.
  \item  Create padded kernel $K^{FFT}_{ij}$.
  \item  Copy $I^{+}_{ij}$ into $I^{FFT}_{ij}$.
  \item  Copy $K_{ij}$ to center of $K^{FFT}_{ij}$.
  \item  In place 2D FFT $I^{FFT}_{ij}$.
  \item  In place 2D FFT $K^{FFT}_{ij}$.
  \item  Multiply $I^{FFT}_{ij}$ by $K^{FFT}_{ij}$ and return to $I^{FFT}_{ij}$.
  \item  In place 2D IFFT $K^{FFT}_{ij}$.
  \item  In place 2D circular Shift $I^{FFT}_{ij}$.
  \item  Multiply $I^{FFT}_{ij}$ sub-region corresponding to $I_{ij}$ by
    $W_{ij}$ and add to $F_{ij}$.
\end{itemize}
    These steps are described in greater detail in Appendix B.
  }
\label{fig:wide_field_flow}
\end{figure}

  Section 2 provided a high level description of using FFTs for 2D
convolution with variable kernels.  This appendix is meant to provide
the more specific details necessary for implementation. 
Figure~\ref{fig:wide_field_flow} presents pictorially the precise
computations, data structures and data movements necessary to execute 2D
convolution algorithm with multiple kernels.  The inputs consist of the
image to be filtered and a grid with a kernel at each node.  The
algorithm proceeds by looping  (in parallel) over each kernel in the
grid $K_{ij}$ and executing the following steps: 
\begin{itemize}
  \item  Determine boundaries of input sub-image $I_{ij}$.
         Using kernel grid determine area affected by kernel $K_{ij}$.
  \item  Compute corresponding weights $W_{ij}$.
         Determine number of kernels contributing to each part
         of the sub-image.  Linearly weight each pixel by its distance
         from the center of kernel.
  \item  Determine boundaries of filtered sub-image $F_{ij}$
         (same as relative position as $I_{ij}$).
  \item  Determine boundaries of sub-image padded by kernel $I^{+}_{ij}$
         by adding N/2 pixel border to $I_{ij}$. 
  \item  Determine size of and create padded sub-image $I^{FFT}_{ij}$
         by padding $I^{+}_{ij}$ out to an optimal FFT size (see Appendix A).
  \item  Create padded kernel $K^{FFT}_{ij}$ (same size as $I^{FFT}_{ij}$).
  \item  Copy $I^{+}_{ij}$ into $I^{FFT}_{ij}$ so that sub-image $I_{ij}$
         is in the center.
  \item  Copy $K_{ij}$ to center of $K^{FFT}_{ij}$.
  \item  In place 2D FFT $I^{FFT}_{ij}$.
  \item  In place 2D FFT $K^{FFT}_{ij}$.
  \item  Multiply $I^{FFT}_{ij}$ by $K^{FFT}_{ij}$ and return to $I^{FFT}_{ij}$.
  \item  In place 2D IFFT $K^{FFT}_{ij}$.
  \item  In place 2D circular Shift $I^{FFT}_{ij}$.  Full half shift in both x and y
         directions.
  \item  Multiply $I^{FFT}_{ij}$ sub-region corresponding to $I_{ij}$ by
         $W_{ij}$ and add to $F_{ij}$.
\end{itemize}

% Appendix with letter and title:
\appendix{C}
\appendixtitle{Dynamic Load Balancing ``Balls into Bins''}

  The classic ``Balls into Bins'' problem has been well studied (see
\cite{Raab1998,Berenbrink2000} and references therein).
In this appendix results are derived for this classic problem as it
pertains to pipeline image processing.  The notation used in this
section is different from the main body of the text, but consistent with
what is used in the wider literature where $M$ is the number of
``balls'' (i.e., the amount work to be done) and $N$ is the number of
``bins'' (i.e., the number of processors that can be used).

  In an image processing context $M$ is the average number of detections
that can be expected in a given image.  In a parallel processing
pipeline, the image will be divided up into $N$ sub-images. Let $\lambda
= M/N$ be the expected number of detections in each sub-image, and
$p_\lambda(m)$ be the probability that that a given sub-image will have
exactly $m$ detections.  Likewise, let $P_\lambda(m) = \sum_{m'=0}^{m}
p_\lambda(m')$ be the probability that a given sub-image has up to $m$
detections in it.

  In most analysis of this type the focus is on the expected maximum
number of ``balls'' that would fall in a given ``bin.''  In a real-time
image processing pipeline environment, it is more typical to work with
the function $M_f$ which is the number detections at a given a failure rate
$f$.  In other words, $f$ is the probability that a given processor in
the system will have more than $M_f$ detections to process.  This
approach is necessary in order to bound the amount of work that is
allowed to during the detections step so as not to miss any real-time
deadlines.  In general, $M_f$ is given by the following equation
  \begin{equation}
      P_\lambda(M_f)^N = 1 - f
  \end{equation}
Finding $f$ for a given value of $M_f$ is simply a matter of evaluating
$P_\lambda$.  Finding $M_f$ for a given $f$ is slightly more difficult
but can be accomplished by using the following coupled recursive equations
\begin{eqnarray}
  N_f^0 & = & 0 \nonumber \\
  P^0   & = & p_\lambda(N_f^0) \nonumber \\
  N_f^{i+1} & = & \{ N_f^i + 1 : P^i < (1 - f)^N \nonumber \\
  P^{i+1} & = & \{ P^i + p_\lambda(N_f^{i+1}) : N_f^{i+1} > N_f^i
\end{eqnarray}
The above formula will work for arbitrary $p_\lambda$, but typically
we are interested in the case where the detections are independent
random events and $p_\lambda$ follows a Poisson distribution
  \begin{equation}
      p_\lambda(m) = \frac{\lambda^m e^{-\lambda}}{m!}
  \end{equation}
In this case, $P_\lambda$ is given by the unnormalized incomplete Gamma
function
  \begin{equation}
      P_\lambda(m) = \tilde{\Gamma}(m+1,\lambda)
                   = \frac{\Gamma(m+1,\lambda)}{m!} ~ ,
  \end{equation}
where $\Gamma$ is the incomplete Gamma function given by the following
sum
  \begin{equation}
     \Gamma(m+1,\lambda) = m! \sum_{m'=0}^m
                         \frac{\lambda^{m'} e^{-\lambda}}{m'!} ~ .
  \end{equation}
Thus, $M_f$ can be computed by inverting the above function
  \begin{equation}
      M_f + 1 = \tilde{\Gamma}^{-1}((1-f)^{1/N},\lambda)
  \end{equation}

\begin{figure}
  \centerline{\includegraphics[width=6.5in]{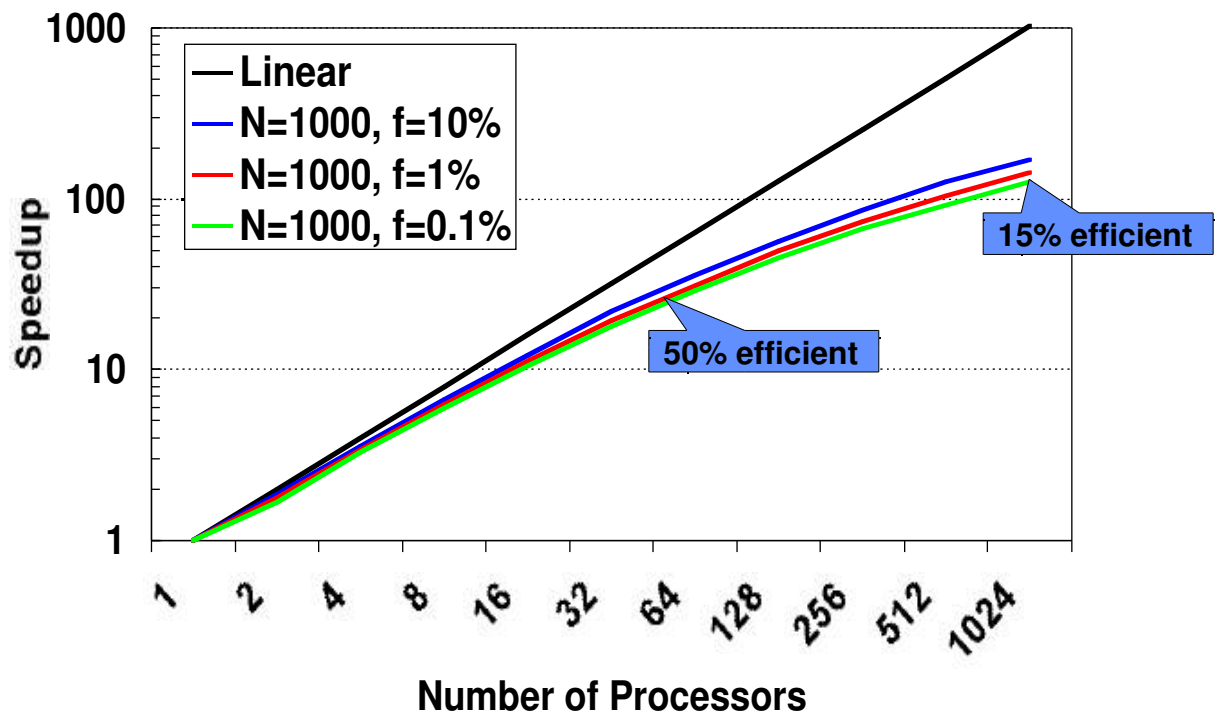}}
  \caption{{\bf Static Speedup.}  Speedup obtainable for M=1000
  tasks (``balls'') and given failure rates using a static
  assignment scheme.  Random fluctuations significantly bound
  speedup.
  }
\label{fig:static_speedup}
\end{figure}

  For a given $M_f$ the speedup on a parallel computer is given by
  \begin{equation}
      {\rm Speedup} = \frac{M}{M_f}
  \end{equation}
If $M_f = \lambda$ then the speedup is linear.  Random
fluctuations in the distribution of detections will mean that some
processors will have more detections resulting in sub-linear speedups. 
Figure~\ref{fig:static_speedup} shows the speedup
on various numbers of processors for different failure rates and
indicates that even for a modest number of processors over half the
processors will be idled due to load imbalances.

\begin{figure}
  \centerline{\includegraphics[width=6.5in]{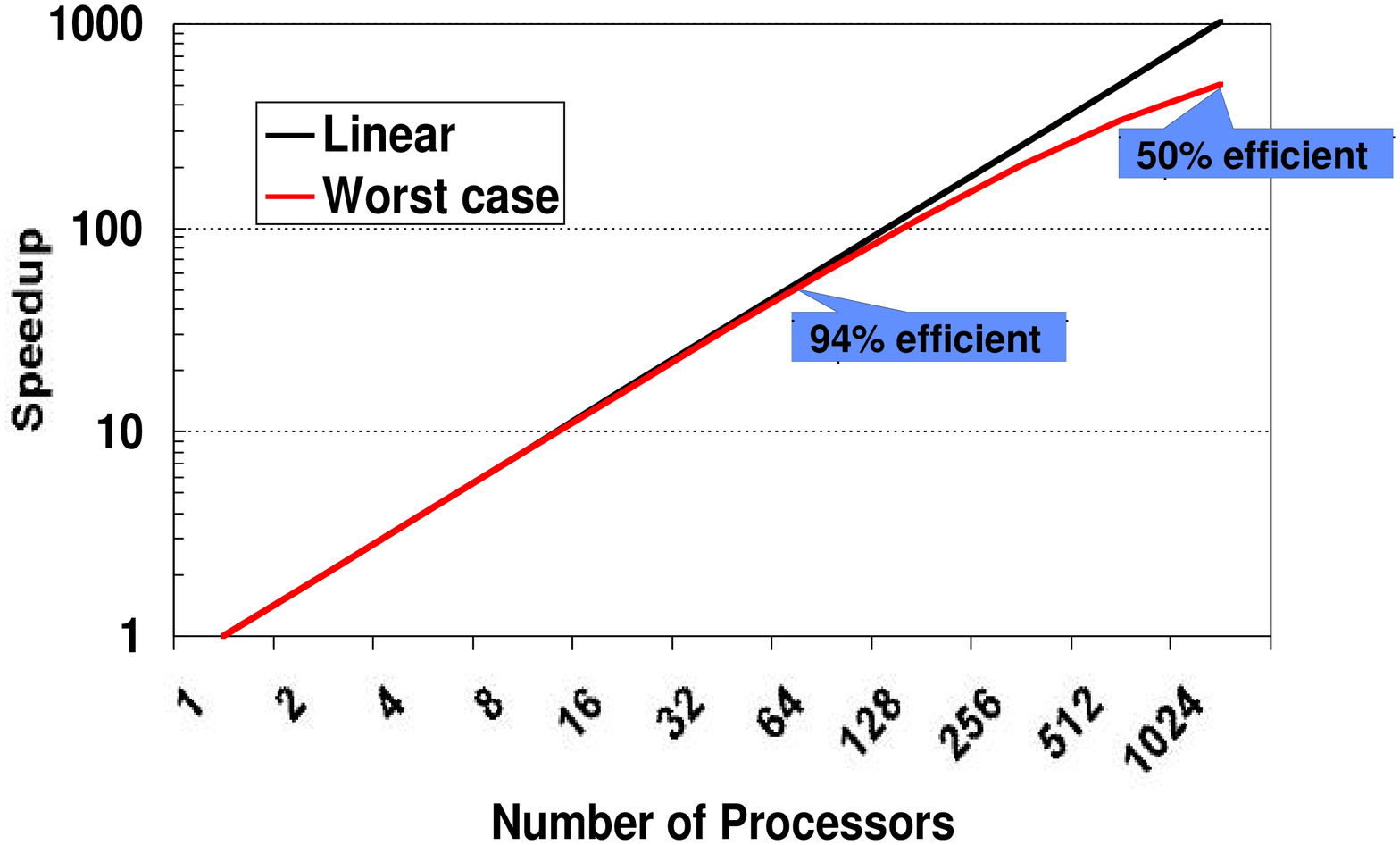}}
  \caption{{\bf Dynamic Speedup.}  Speedup obtainable for M=1000
   tasks and f = 0 using a dynamic work assignment scheme.  This
   scheme provides good speedups even in the worst case.
  }
\label{fig:dynamic_speedup}
\end{figure}

  Alleviating the effects of random work loads requires adopting
dynamic schemes that allow work to be assigned to processors as
they become available.  Assuming the work can be broken up with
a granularity $g$ (i.e. the smallest amount of work is $g M$),
then the worst case speedup ($f = 0$) is given by
  \begin{equation}
    {\rm Speedup} = \frac{M}{\lambda + g M}
                  = \frac{M}{M/N + g M}
                  = \frac{N}{1 + g N} ~ .
  \end{equation}
Asymptotically, for large $N$ the speedup converges to $1/g$.
Figure~\ref{fig:dynamic_speedup} shows the speedup obtainable
using this type of strategy, which is significantly better than
what can be obtained in the static situation.

%% not optional:

\newpage

%%%%%%%%%%%%%%%%%%%%%%%%%%%%%%%%%%%%%%%%%%%%%%%%%%%%%%%%%%%%%%%%%%%%%%%
% TABLES
%%%%%%%%%%%%%%%%%%%%%%%%%%%%%%%%%%%%%%%%%%%%%%%%%%%%%%%%%%%%%%%%%%%%%%%

\begin{table}[tbh]
\begin{tabular}{cccccc}
\hline
Parallel    & Compute Latency & Communication & Load      & Software \\
Approach    & Overhead        & Overhead      & Balancing & Overhead \\
\hline
Image       & high            & high          & excellent & application \\
Kernel      & medium          & low           & good      & math library \\
Row/Column  & low             & high          & good      & math kernel  \\
Instruction & low             & unknown       & excellent & OS/hardware  \\
\hline
\end{tabular}
\caption{ {\bf Parallel Overheads.}
  Computation, communication, load balancing and coding overheads for
exploiting different levels of parallelism for doing a 2D convolution.
}
\label{tab:parallel_overheads}
\end{table}

\begin{table}[tbh]
\begin{tabular}{ccccc}
\hline
Image size & Kernel grid & Number of  & Measured & Parallel \\
   $M$     &    $N_K$    & Processors & Speedup  & Efficiency (\%) \\
\hline
% R = 20*(8*512)^2*18 / (8*512)^2 + 4*1024^2 = 288
   1024    &     8       &     1      & 1.00     & 100 \\
   1024    &     8       &     2      & 1.96     &  98 \\
   1024    &     8       &     4      & 3.88     &  97 \\
   1024    &     8       &     8      & 7.76     &  97 \\
   1024    &     8       &     16     & 15.2     &  95 \\
   1024    &     8       &     32     & 29.0     &  91 \\
\hline
% R = 20*(16*512)^2*18 / (16*512)^2 + 4*2048^2 = 288
   2048    &     16      &     1      & 1.00     & 100 \\
   2048    &     16      &     2      & 1.93     &  97 \\
   2048    &     16      &     4      & 3.86     &  97 \\
   2048    &     16      &     8      & 7.79     &  97 \\
   2048    &     16      &     16     & 15.3     &  97 \\
   2048    &     16      &     32     & 30.1     &  94 \\
\hline
% R = 20*(32*512)^2*18 / (32*512)^2 + 4*4096^2 = 288
   4096    &     32      &     1      & 1.00     & 100 \\
   4096    &     32      &     2      & 1.95     &  98 \\
   4096    &     32      &     4      & 3.80     &  95 \\
   4096    &     32      &     8      & 7.73     &  97 \\
   4096    &     32      &     16     & 15.1     &  94 \\
   4096    &     32      &     32     & 30.1     &  94 \\
\hline
\end{tabular}
\caption{ {\bf Parallel speedups.}
Measured parallel speedup and parallel efficiency obtained for different
image sizes and different numbers of processors.  In each case the
kernel size was $N = 100$.
}
\label{tab:parallel_speedups}
\end{table}

\clearpage
\newpage

%%%%%%%%%%%%%%%%%%%%%%%%%%%%%%%%%%%%%%%%%%%
% FIGURE CAPTIONS
%%%%%%%%%%%%%%%%%%%%%%%%%%%%%%%%%%%%%%%%%%%

%%%%%%%%%%%%%%%%%%%%%%%%%%%%%%%%%%%%%%%%%%%%%%%%%%%%%%%%%%%%%%%%%%%%%%%
% Text Figures.
%%%%%%%%%%%%%%%%%%%%%%%%%%%%%%%%%%%%%%%%%%%%%%%%%%%%%%%%%%%%%%%%%%%%%%%

%%%%%%%%%%%%%%%%%%%%%%%%%%%%%%%%%%%%%%%%%%%%%%%%%%%%%%%%%%%%%%%%%%%%%%%
% Appendix A Figures.
%%%%%%%%%%%%%%%%%%%%%%%%%%%%%%%%%%%%%%%%%%%%%%%%%%%%%%%%%%%%%%%%%%%%%%%

%%%%%%%%%%%%%%%%%%%%%%%%%%%%%%%%%%%%%%%%%%%%%%%%%%%%%%%%%%%%%%%%%%%%%%%
% Appendix B Figures.
%%%%%%%%%%%%%%%%%%%%%%%%%%%%%%%%%%%%%%%%%%%%%%%%%%%%%%%%%%%%%%%%%%%%%%%

%%%%%%%%%%%%%%%%%%%%%%%%%%%%%%%%%%%%%%%%%%%%%%%%%%%%%%%%%%%%%%%%%%%%%%%
% Appendix C Figures.
%%%%%%%%%%%%%%%%%%%%%%%%%%%%%%%%%%%%%%%%%%%%%%%%%%%%%%%%%%%%%%%%%%%%%%%

%% This command is necessary! ==>>
\end{article}

\begin{references}

\bibitem{Berenbrink2000}
         P. Berenbrink, A. Czumaj, A. Steger and B. Vocking,
         Balanced Allocations: The Heavily Loaded Case,
         in ``Proceedings of the 32nd Annual ACM Symposium on Theory of Computing,''
         pp 745-754, Portland, OR, May 21-23 2000.  ACM Press, New York, NY.
\bibitem{Frigo1998}
         M. Frigo and S. Johnson,
         FFTW: An Adaptive Software Architecture for the FFT,
         ICASSP Proceedings (1998), vol 3., p 1381.
\bibitem{IDL}
         Interactive Data Language by Research Systems, Inc.
         Boulder, CO
         http://www.rsi.com/
\bibitem{Kepner1998}
         J. Kepner, M. Gokhale, R. Minnich, A. Marks and J. DeGood,
         Interfacing Interpreted and Compiled Languages to Support
         Applications on a Massively Parallel Network of
         Workstations (MP-NOW), Cluster Computing 2000, volume 3, number 1, page 66
\bibitem{Kepner2000a}
         J. Kepner
         Integration of VSIPL and OpenMP into a Parallel Image
         Processing Environment, J. Kepner, proceedings of the the
         fourth High Performance Embedded Computing Workshop
         (HPEC 2000), September 20-22, 2000, MIT Lincoln
         Laboratory, Lexington, MA 
\bibitem{Kepner2000b}
         J. Kepner,
         Exploiting VSIPL and OpenMP for Parallel Image Processing,
         in ``Proceedings of ADASS X'' (editors), Boston, MA 2000
\bibitem{Matson1999}
        T. Mattson and R. Eigenmann,
        Parallel Programming with OpenMP,
        Supercomputing'99, Nov 13, 1999, Orlando, FL
\bibitem{Moler2001}
         C. Moler and S. Eddins,
         Faster Finite Fourier Transforms,
         Matlab News \& Notes, Winter 2001
\bibitem{OpenMP}
         OpenMP: Simple, Portable, Scalable Programming,
         http://www.openmp.org/
\bibitem{Raab1998}
         M. Raab \&  A. Steger,
         Bins into Balls --- A Simple and Tight Analysis
         In 2nd International Workshop on Randomization and Approximation Techniques in Computer Science
         (RANDOM'98), pages 159--170, 1998.
\bibitem{Shirazi1995}
         B. A. Shirazi, A R. Hurson, K. M. Kavi,
         Scheduling and Load Balancing in Parallel and Distributed Systems,
         IEEE Computer Society Press, 1995
\bibitem{Stockham1966}
          T. G. Stockham, High Speed Convolution and Correlation,
          Spring Joint Computer Conference, AFIPS Proceedings, 28, pp 229-233, 1966
\bibitem{VSIPL}
         Vector, Signal, and Image Processing Library
         http://www.vsipl.org/

\end{references}
\end{document}